\documentclass[alpha-refs]{wiley-article}

\usepackage{graphicx}
\usepackage{longtable}
\usepackage{tabulary}
\usepackage{booktabs,array,multirow}
\usepackage{amsfonts,amsmath}
\usepackage{natbib}
\usepackage{url}
\usepackage{hyperref}
\hypersetup{colorlinks=false,pdfborder={0 0 0}}
\usepackage[utf8]{inputenc}
\usepackage[english]{babel}

\newcommand{\mref}[1]{(\ref{#1})}

\newcommand{\RR}{\mathbb{R}}
\newcommand{\dd}{\,d}

\newcommand{\sumi}{\sum_{i = 1}^{N}}

\newcommand{\E}{\operatorname{E}}
\newcommand{\BIC}{\mathrm{BIC}}

\newcommand{\Prob}{\operatorname{P}}

\newcommand{\given}{\operatorname{|}}
\newcommand{\X}{\mathbf{X}}
\newcommand{\M}{\mathbf{M}}
\newcommand{\G}{\mathcal{G}}
\newcommand{\D}{\mathcal{D}}

\newcommand{\PXi}{\Pi_{X_i}}
\newcommand{\XPi}{X_i \given \PXi}
\newcommand{\T}{\Theta_{X_i}}
\newcommand{\TT}{\T \given \PXi}
\newcommand{\piijk}{\pi_{ik \given j}}
\newcommand{\dXi}{\delta_{X_i}}
\newcommand{\DXi}{\Delta_{X_i}}
\newcommand{\GXi}{\Gamma_{X_i}}

\newcommand{\U}{\mathbf{U}}
\newcommand{\Ut}{\U^{(t)}}
\newcommand{\Xt}{\X^{(t)}}
\newcommand{\Xtt}{\X^{(t - 1)}}
\newcommand{\Xit}{X_i^{(t)}}
\newcommand{\PXit}{\Pi_{X_i^{(t)}}}
\newcommand{\Z}{\mathbf{Z}}
\newcommand{\Zt}{\Z^{(t)}}
\newcommand{\Ztt}{\Z^{(t - 1)}}
\newcommand{\Zjt}{Z_j^{(t)}}
\newcommand{\PZjt}{\Pi_{Z_j^{(t)}}}

\corraddress{Marco Scutari,
  Istituto Dalle Molle di Studi sull'Intelligenza Artificiale (IDSIA),
  Galleria 2, Via Cantonale 2c,
  6928 Manno, Switzerland}
\corremail{scutari@idsia.ch}
\papertype{Original Article}

\title{Bayesian Network Models for Incomplete and Dynamic Data}
\author[1]{Marco Scutari}
\affil[1]{Istituto Dalle Molle di Studi sull'Intelligenza Artificiale (IDSIA)}
\runningauthor{Marco Scutari}

\begin{document}

\maketitle
\selectlanguage{english}
\begin{abstract}

Bayesian networks are a versatile and powerful tool to model complex phenomena
and the interplay of their components in a probabilistically principled way.
Moving beyond the comparatively simple case of completely observed, static data,
which has received the most attention in the literature, in this paper we will
review how Bayesian networks can model dynamic data and data with incomplete
observations. Such data are the norm at the forefront of research and
in practical applications, and Bayesian networks are uniquely positioned to
model them due to their explainability and interpretability.

\end{abstract}

\section{Introduction}
\label{sec:intro}

Bayesian networks (BNs) are a probabilistic graphical model that is ideally
suited to study complex phenomena through the interactions of their components.
By representing the latter as nodes in a graph, and connecting them with arcs
that encode how those components interact with each other, they provide a
high-level qualitative view of the phenomenon under investigation built on
strong theoretical foundations.

BNs were initially developed in the 1980s in the context of expert systems, with
the seminal work summarised in \citet{pearl} and \citet{neapolitan-old}. Later
monographs by \citet{castillo}, \citet{neapolitan}, \citet{cowell-book},
\citet{korb} provide extensive coverage of the properties of BNs and their use;
with \citet{koller} being the most complete reference book up to date. In that
context, ``expert systems'' were conceived as the combination of a formal
representation of domain-specific knowledge gathered from subject matter experts
and an ``inference engine'' that could use that representation to answer
arbitrary queries. BNs can represent complex phenomena in a modular way due the
conditional independence assumptions they encode, allowing them to scale
beyond trivial problems and to be used to develop efficient inference
algorithms.

As noted in \citet{cowell-book}, by the end of the 1990s the limitations of
constructing expert systems had become more and more apparent and were pushing
research towards learning the graphical structure and the parameters of BNs from
data. For many phenomena that were and have since been of interest in the
scientific community, expert systems are too complex for the modeller to be able
to elicit the required domain knowledge from experts, even assuming expert
knowledge is available in the first place (or with sufficient consensus to be
trusted). However, at the same time the availability of increasingly large data
sets made it more attractive to use them as an alternative source of information
to construct BNs.

Given the limited computational power available at the time, this transition
from expert elicitation to learning from data was made possible by making a
number of simplifying assumptions on the nature of the data \citep{heckerman}.
Two assumptions in particular have been made in the vast majority of the
literature since then. Firstly, that data are \emph{complete}; that is, that
all variables are completely observed for all samples. And secondly, that
samples are \emph{independent} from each other; which specifically excludes
dynamic data such as time series in which samples are associated with each
other.

However, problems at the forefront of research and in practical applications
have only been increasing in complexity, making these assumptions too strong in
many settings. In this paper, we will review how BNs can be used to model
dynamic data and data with incomplete observations, providing several examples
in which they have been used successfully to develop cutting-edge applications.
The comparatively simple case of complete, static data has certainly received
the most attention in the literature, for historical reasons and for its
probabilistic and computational simplicity; but BNs are by no means limited to
modelling only such data.

The contents of this review are organised as follows. In Section
\ref{sec:background} we introduce to BNs in a general setting, including their
underlying distributional assumptions and available learning approaches. In
Section \ref{sec:dynamic} we discuss BNs targeted at dynamic (that is, temporal)
data, starting from their definition and main properties (Section
\ref{sec:dbn-intro}) and then showing them to be a general approach that
subsumes several other models in the literature (Section
\ref{sec:special-cases}). In Section \ref{sec:incomplete} we discuss incomplete
data and how they can be correctly used for parameter (Section
\ref{sec:paramlearn}) and structure learning. Finally, in Section
\ref{sec:applications} we provide an overview of available software
implementations (Section \ref{sec:software}) and real-world applications
(Section \ref{sec:notable}) of the BNs we discussed.

\section{Background and Notation}
\label{sec:background}

BNs are a class of graphical models in which the nodes of a directed acyclic
graph (DAG) $\G$ represent a set $\mathbf{X} = \{X_1, \ldots, X_N\}$ of random
variables describing some quantities of interest. The arcs connecting those
nodes express direct dependence relationships, with graphical separation in $\G$
implying conditional independence in probability. As a result, $\G$ induces the
factorisation
\begin{equation}
  \Prob(\X \given \G, \Theta) = \prod_{i=1}^N \Prob\left(\XPi, \T\right),
\label{eq:parents}
\end{equation}
in which the joint probability distribution of $\X$ (with parameters $\Theta$)
decomposes into one \emph{local distribution} for each $X_i$ (with parameters
$\T$, $\bigcup_{X_i \in \X} \T = \Theta$) conditional on its parents $\PXi$.
Assuming $\G$ is sparse\footnote{There is no universally accepted threshold on
the number of arcs for a DAG to be called ``sparse''; typically it is taken to
have $O(cN)$ arcs, with $c$ between $1$ and $5$.}, BNs provide a compact
representation of both low- and high-dimensional probability distributions.

BNs are also very flexible in terms of distributional assumptions; but while in
principle we could choose any probability distribution for $\X$, the literature
has mostly focused on three cases for analytical and computational reasons.
\emph{Discrete BNs} \citep{heckerman} assume that both $\X$ and the $X_i$ are
multinomial random variables. Local distributions take the form
\begin{align}
  &\XPi \sim \mathit{Mul}\left(\piijk\right),&
  &\piijk = \Prob\left(X_i = k \given \PXi = j\right);
\label{eq:cpt}
\end{align}
their parameters $\piijk$ are the conditional probabilities of $X_i$ given each
configuration of the values of its parents, usually represented as a conditional
probability table for each $X_i$. \emph{Gaussian BNs}
\citep[GBNs;][]{heckerman3} model $\X$ with a multivariate normal random
variable and assume that the $X_i$ are univariate normals linked by linear
dependencies. The parameters of the local distributions can be equivalently
written \citep{weatherburn} as the partial correlations $\rho_{X_i, X_j \given
\PXi \setminus X_j}$ between $X_i$ and each parent $X_j$ given the other
parents; or as the coefficients $\boldsymbol{\beta}_{X_i}$ of the linear
regression model
\begin{align}
  &X_i = \mu_{X_i} + \PXi\boldsymbol{\beta}_{X_i} + \varepsilon_{X_i},&
  &\varepsilon_{X_i} \sim N\left(0, \sigma^2_{X_i}\right),
\label{eq:reg}
\end{align}
so that $\XPi \sim N\left(\mu_{X_i} + \PXi\boldsymbol{\beta}_{X_i},
\sigma^2_{X_i}\right)$.  Finally, \emph{conditional linear Gaussian BNs}
\citep[CLGBNs;][]{clgbn} combine discrete and continuous random
variables in a mixture model:
\begin{itemize}
  \item discrete $X_i$ are only allowed to have discrete parents (denoted
    $\DXi$), and are assumed to follow a multinomial distribution as in
    \mref{eq:cpt};
  \item continuous $X_i$ are allowed to have both discrete and continuous
    parents (denoted $\GXi$, $\DXi \cup \GXi = \PXi$), and their local
    distributions are
    \begin{equation*}
      \XPi \sim N\left(\mu_{X_i, \dXi} +
                  \Gamma_{X_i}\boldsymbol{\beta}_{X_i, \dXi},
                    \sigma^2_{X_i, \dXi}\right)
    \end{equation*}
    which can be written as a mixture of linear regressions
    \begin{align*}
      &X_i = \mu_{X_i, \dXi} + \Gamma_{X_i}\boldsymbol{\beta}_{X_i, \dXi} +
              \varepsilon_{X_i, \dXi},&
      &\varepsilon_{X_i, \dXi} \sim N\left(0, \sigma^2_{X_i, \dXi}\right)
    \end{align*}
    against the continuous parents with one component for each configuration
    $\dXi$ of the discrete parents $\DXi$. If $X_i$ has no discrete parents,
    the mixture reverts to a single linear regression like that in \mref{eq:reg}.
\end{itemize}

The task of learning a BN from a data set $\D$ containing $n$ observations is
performed in two steps:
\begin{equation*}
  \underbrace{\Prob(\G, \Theta \given \D)}_{\text{learning}} =
    \underbrace{\Prob(\G \given \D)}_{\text{structure learning}} \cdot
    \underbrace{\Prob(\Theta \given \G, \D)}_{\text{parameter learning}}.
\end{equation*}
\emph{Structure learning} consists in finding the DAG $\G$ that encodes the
dependence structure of the data, thus maximising $\Prob(\G \given \D)$ or
some alternative goodness-of-fit measure; \emph{parameter learning} consists
in estimating the parameters $\Theta$ given the $\G$ obtained from structure
learning. Both steps can integrate data with expert knowledge through the use
of suitable prior distributions on $\G$ and $\Theta$ \citep[see for
example][]{csprior,mukherjee,linda}. If we assume that parameters in different
local distributions are independent and that the data contain no missing values
\citep{heckerman}, we can perform parameter learning independently for each
$X_i$ because following \mref{eq:parents}
\begin{equation*}
  \Prob(\Theta \given \G, \D) = \prod_{i=1}^N \Prob\left(\TT, \D\right).
\end{equation*}
Furthermore, assuming $\G$ is sparse, each local distribution $\XPi$ will
involve only a few variables and thus will have a low-dimensional parameter
space, making parameter learning computationally efficient.

On the other hand, structure learning is well known to be both NP-hard
\citep{nphard} and NP-complete \citep{npcomp}, even under unrealistically
favourable conditions such as the availability of an independence and inference
oracle \citep{nplarge}\footnote{Interestingly, some relaxations of BN structure
learning are not NP-hard; see for example \citet{notnphard} on learning the
structure of causal networks.}. This is despite the fact that if we take
\begin{equation}
  \Prob(\G \given \D) \propto \Prob(\G)\Prob(\D \given \G),
\label{eq:learning}
\end{equation}
following \mref{eq:parents} we can decompose the marginal likelihood
$\Prob(\D \given \G)$ into one component for each local distribution
\begin{equation}
    \Prob(\D \given \G)
    = \int \Prob(\D \given \G, \Theta) \Prob(\Theta \given \G) \dd\Theta
    = \prod_{i=1}^N \int \Prob\left(\XPi, \T\right) \Prob\left(\TT\right) \dd\T;
\label{eq:decomp}
\end{equation}
and despite the fact that each component can be written in closed form for
discrete BNs \citep{heckerman}, GBNs \citep{heckerman3} and CLGBNs
\citep{bottcher}. The same is true if we replace $\Prob(\D \given \G)$ with
frequentist goodness-of-fit scores such as BIC \citep{schwarz}, which is
commonly used in structure learning because of its simple expression:
\begin{equation*}
  \BIC(\G, \Theta \given \D) =
    \sumi \log \Prob\left(\XPi, \T\right) - \frac{\log(n)}{2} \left|\T\right|.
\end{equation*}
Compared to marginal likelihoods, BIC also has the advantage that it does not
depend on any hyperparameter, while converging to $\log\Prob(\D \given \G)$ as
$n \to \infty$. These score functions have two important properties:
\begin{itemize}
  \item they allow local computations because, following \mref{eq:parents}, they
    \emph{decompose} into one component for each local distribution;
  \item they take the same value for all the DAGs that encode the same
    probability distribution (\emph{score equivalence}), which can then be
    grouped in \emph{equivalence classes} \citep{chickering}.\footnote{All DAGs
    in the same equivalence class have the same underlying undirected graph and
    v-structures (patterns of arcs like $X_i \rightarrow X_j \leftarrow X_k$,
    with no arcs between $X_i$ and $X_k$).}
\end{itemize}
Structure learning via score maximisation is usually based on general-purpose
heuristic optimisation algorithms, adapted to take advantage of these two
properties to increase the speed of structure learning \citep{stco17}. The most
common are \emph{greedy search} strategies such as hill-climbing and tabu search
\citep{norvig} that employ local moves designed to affect only one or two local
distributions in each iteration; other options explored in the literature
include genetic algorithms \citep{genetic} and ant colony optimisation
\citep{ant}. Learning equivalence classes directly (as opposed to DAGs) can be
done along the same lines with the Greedy Equivalence Search \citep[GES;][]{ges}
algorithm. Exact maximisation of $\Prob(\D \given \G)$ and BIC has also become
feasible in recent years thanks to increasingly efficient pruning of the space
of DAGs and tight bounds on the scores \citep{cutting,suzuki17,scanagatta}.

Another option for structure learning is using conditional independence tests to
learn conditional independence constraints from $\D$, and thus to identify which
arcs should be included in $\G$. The resulting algorithms are called
\emph{constraint-based algorithms}, as opposed to the \emph{score-based
algorithms} we introduced in the previous paragraph; for an overview and a
comparison of these two approaches see \citet{pgm18}. \citet{nplarge} proved
that constraint-based algorithms are also NP-hard for unrestricted DAGs; and
they are in fact equivalent to score-based algorithms given a fixed topological
ordering of the nodes in $\G$ when independence constraints are assessed with
statistical tests related to cross-entropy \citep{cowell}.

Finally, once both $\G$ and $\Theta$ have been learned, we can answer
\emph{queries} about our quantities of interest using the resulting BN as our
model of the world. Common types are \emph{conditional probability} queries, in
which we compute the posterior probability of some variables given evidence on
others; and \emph{most probable explanation} queries, in which we identify the
configuration of values of some variables that has the highest posterior
probability given the values of some other variables. The latter is especially
suited to implement both prediction and imputation of missing data. These
queries can be automated, for any given BN, using either \emph{exact} or
\emph{approximate} inference algorithms that work directly on the BN without the
need for any manual calculation; for an overview of such algorithms see
\citet{crc13}.

Given their ability to concisely but meaningfully represent the world, to
automatically answer arbitrary queries, and to combine data and expert knowledge
in the learning process, BNs can model a wide variety of phenomena effectively.
However, their applicability is not always apparent to practitioners in other
fields due to the strong focus of the literature on the simple scenario in which
data are \emph{static} (as opposed to \emph{dynamic}, that is, with a time
dimension) and \emph{complete} (as in, completely observed). Dynamic data are
central to a number of cutting-edge applications and research in fields as
different as genetics and robotics; and incomplete data are a fact of life in
almost any real-world data analysis. BNs can handle both in a
rigorous way, as we will see in the following.

\section{Dynamic Bayesian Networks}
\label{sec:dynamic}

Dynamic BNs (DBNs\footnote{Confusingly, discrete BNs are sometimes called DBNs
to be consistent with Gaussian BNs being called GBNs.}) combine classic (static)
BNs and Markov processes to model dynamic data in which each individual is
measured repeatedly over time, such as longitudinal or panel data.  An
approachable introduction is provided for instance in \citet{murphy}. They have
major applications in engineering \citep{pavlovic,security}, medicine
\citep{sem-ecr}, genetics and systems biology \citep{perrin}. The term
``dynamic'' in this context implies we are modelling a dynamic system, not
necessarily that the network changes over time.

\subsection{Definitions and Properties}
\label{sec:dbn-intro}

For simplicity, let's assume at first that we are operating in discrete time:
our system consists of one set $\X^{(t)}$ of random variables for each of
$t = 1, \ldots, T$ time points. We can model it as a DBN with a Markov process
of the form
\begin{equation}
  \Prob\left(\X^{(0)}, \ldots, \X^{(T)}\right) =
    \Prob\left(\X^{(0)}\right)
    \prod_{t = 1}^T \Prob\left(\Xt \given \Xtt\right).
\label{eq:dbn-def}
\end{equation}
where $\Prob(\X^{(0)})$ gives the initial state of the process and
$\Prob(\X^{(t)} \given \X^{(t - 1)})$ defines the transition between times $t -
1$ and $t$. We can model this transition with a 2-time BN (2TBN) defined over
$(\X^{(t - 1)}, \X^{(t)})$, in which we naturally assume that any arc between a
node in $t - 1$ and a node in $t$ must necessarily be directed towards the
latter following the arrow of time. When modelling $\X^{(t)}$, the nodes in
$\X^{(t - 1)}$ only appear in the conditioning; we take them to be essentially
fixed and to have no free parameters, so we leave them as root nodes. After all,
$\X^{(t - 1)}$ will be stochastic in $\Prob(\X^{(t - 1)} \given \X^{(t - 2)})$
and it would not be consistent with \mref{eq:dbn-def} to treat $\X^{(t - 1)}$ as
a stochastic quantity twice! Then, following \mref{eq:parents} we can write
\begin{equation} \Prob\left(\Xt \given \Xtt\right) = \prod_{i = 1}^N
\Prob\left(\Xit \given \PXit\right), \end{equation} and we usually assume that
the parameters associated with the local distributions do not change over time
to make the process time-homogeneous.

\begin{figure}
\begin{center}
  \includegraphics[width=0.8\textwidth]{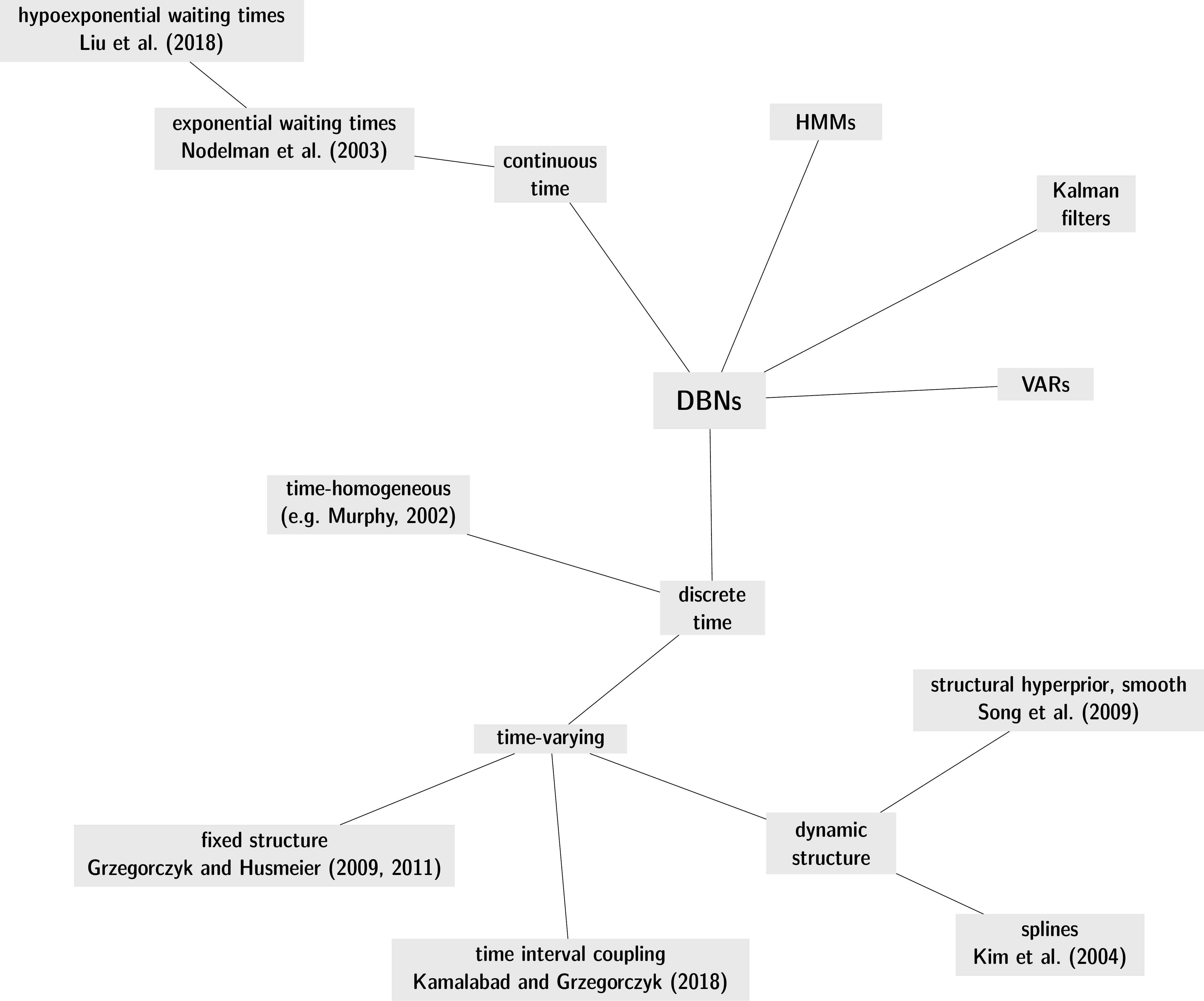}
  \caption{Diagram illustrating the relationships between the papers and the
  approaches covered in Section \ref{sec:dynamic}.}
  \label{fig:dynamic-mindmap}
\end{center}
\end{figure}

These choices are motivated by computational and statistical simplicity: there
is no intrinsic limitation in the construction of DBNs that prevents them from
modelling dependencies that stretch further back in time or, for that matter,
trends or seasonality. (See Figure \ref{fig:dynamic-mindmap} for an overview of
the models we will discuss in detail below.) \citet{husmeier-nips,husmeier}, for
instance, constructed non-stationary, non-homogeneous DBNs for modelling
continuous data using change-points to capture different regimes in different
time intervals; they balanced this increased flexibility by requiring $\G$ to be
the same for all regimes, allowing only the parameters of the local
distributions to change. This restriction was later relaxed in
\citet{husmeier-latest} by allowing both $\G$ and the parameters to vary but
coupling them between adjacent time intervals. The hyperparameters controlling
the coupling were allowed to vary between segments, following an inverse Gamma
hyperprior. \citet{robinson} also defined a non-stationary DBN with
change-points and associated arc changesets, with truncated geometric priors on
the size of the changesets, number of and interval between change-points; and
\citet{song} constructed DBNs that vary smoothly (not piecewise) over time in
both structure and parameters. \citet{imoto} introduced an even more flexible
model that used spline regression with B-splines to identify $\PXit$. Augmenting
temporal (panel) data with non-temporal (cross-sectional) data for learning DBNs
has been explored as well by \citet{lahdesmaki} using an score that approximates
the resulting intractable likelihood.

Modelling DBNs as discrete time processes is likewise a choice motivated by
mathematical simplicity. \citet{nodelman} originally proposed a class of
continuous-time DBNs (CTBNs) with independent exponential waiting times and
discrete nodes; they are uniquely identifiable since all arcs are
non-instantaneous, and they have a closed-form marginal likelihood as well. More
recently, this work has been expanded in \citet{stella} by replacing exponential
waiting times with hypoexponentials to better reflect the behaviour of data in
several domains. Discrete-time DBNs are certainly simpler than any of these
CTBNs, but that mathematical simplicity comes with important practical
consequences. Firstly, in order to work in discrete time we must choose a
uniform time step (the length of time between $t - 1$ and $t$) for the whole
DBN; but in many real-world phenomena different variables can have very
different time granularities, and those time granularities may vary as well in
the course of data collection, making any single choice for the time step
inappropriate. Secondly, the choice of the time step may obscure the dynamics of
the phenomenon. The implication of using discrete time is that we aggregate all
the state changes in the DBN over the entire course of each time step.  On the
one hand, if variables evolve at slower pace than the time step we are forced to
model the DBN as a higher-order Markov process\footnote{A stochastic process is
a Markov process of order $L$ if $\Xt$ depends only on $\Xtt, \ldots \X^{(t -
L)}$ and is independent from $\X^{(t - L - 1)}, \ldots, \X^{(0)}$. The higher
the order, the further back in time the dependencies can reach.}, resulting in a
much more complex model. On the other hand, if variables evolve at a faster pace
than time step, this averaging will effectively hide state changes and their
interplay into a single summary statistic; hence the DBN will provide a very
poor approximation of the underlying phenomenon. Furthermore, if we are unable
to correctly identify the $\PXit$, $\Xit$ may end up being recursively linked to
the parents of the $\PXit$\footnote{This issue is often called
\emph{entanglement}.}, resulting into dense DBNs that are much more complex than
the real underlying phenomenon and that are difficult to learn from limited
data.

A partial solution to the latter problem is to allow the parents $\PXit$ to be
either in the same time or in the previous time point, modelling the DBN as a
first-order Markov process. When observations represent average or aggregate
measurements over a period of time (say, the $t$th time point corresponds
to the $t$th week's worth of data), it makes sense to allow \emph{instantaneous
dependencies} between variables in the same time point, since the
instantaneousness of the dependence is just a fiction arising from our model
definition. On the other hand, when observations actually correspond to
instantaneous measurements (such as from synchronised sensors) only
\emph{non-instantaneous dependencies} are usually allowed in the model, on the
grounds that conditioning event ($\PXit = \pi_{\Xit}$) should precede in time
the conditioned event ($\Xit = x_{i}^{(t)}$). From a causal perspective, we can
similarly argue that each of the $\PXit$ can only cause $\Xit$ if it precedes
$\Xit$ in time; if that $\PXit$ is in the same time point as $\Xit$ then what we
are modelling is co-occurrence and not causation. This the core idea of
\emph{Granger causality} \citep{granger}, which states that one time series
(such as the $\PXit$) can be said to have a causal influence on a second time
series (such as the $\Xit$) if and only if incorporating past knowledge about
the former improves predictive accuracy for the latter. Therefore, allowing
instantaneous dependencies makes causal reasoning on the DBN markedly more
difficult. The same is true for learning the structure of the DBN in the first
place: learning a general BN is NP-hard \citep{nphard,nplarge}, while learning a
DBN containing only non-instantaneous dependencies is not \citep{notnphard}.
Intuitively, the space of the possible DBNs is much smaller if we do not allow
instantaneous dependencies because there are fewer candidate arcs that we can
include, and because their directions are fixed to follow the arrow of time and
Granger causality\footnote{Interestingly, when the number of available time
points is small, inference based on DBNs is more accurate than methods directly
based on Granger causality; but the opposite is true for longer time series
\citep{granger-dbn}.}.

\subsection{Models That Can Be Expressed as DBNs}
\label{sec:special-cases}

\begin{figure}
  \begin{center}
  \includegraphics[width=0.8\textwidth]{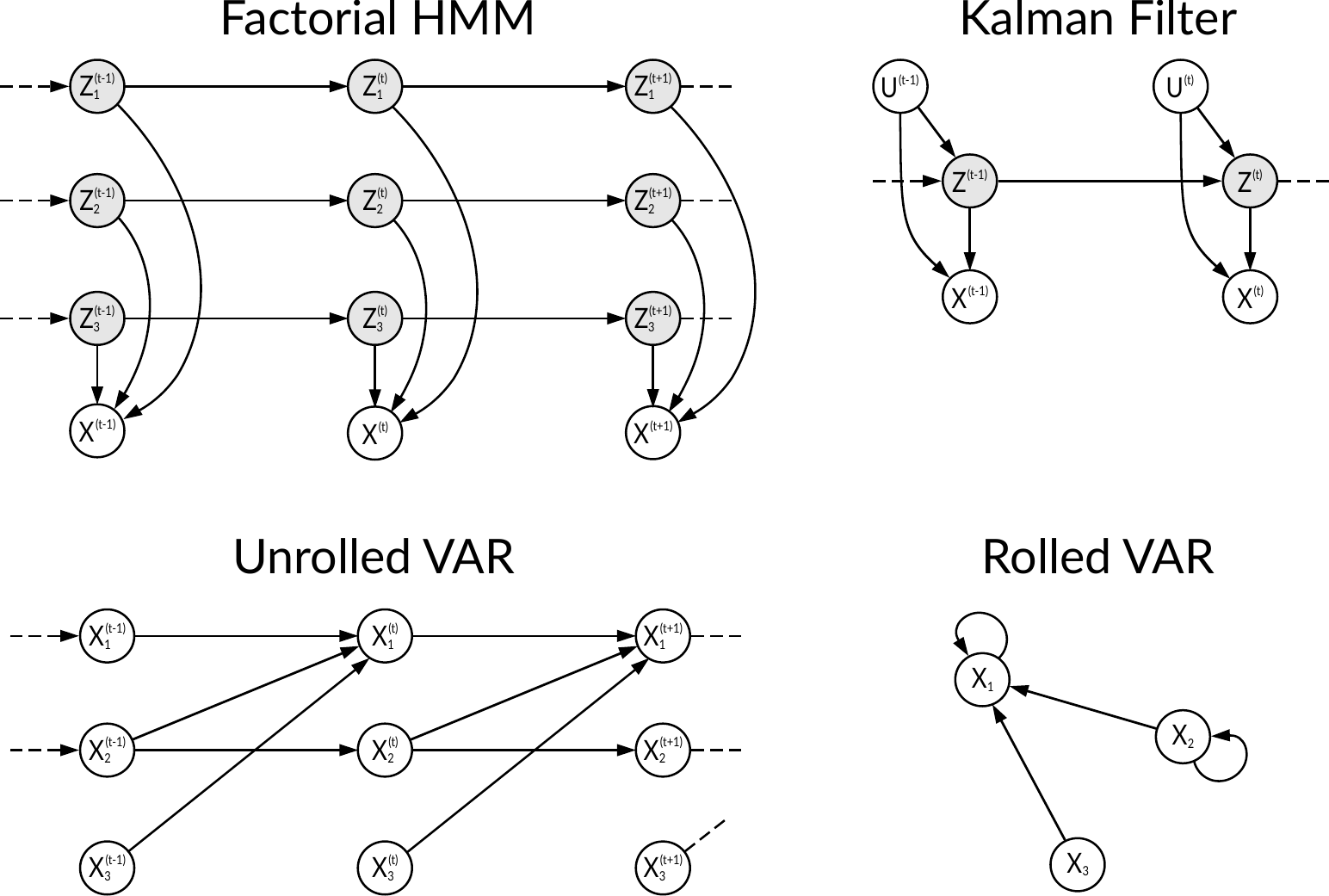}
  \caption{Dynamic models represented as BNs: the factorial HMM from
    \citet[top left]{factorial-hmms}, a simple Kalman filter model from
    \citet[top right]{murphy}, and a VAR model of order 1 in both unrolled
    (bottom left) and rolled forms (bottom right). Shaded nodes correspond
    to latent variables.}
  \label{fig:dbns}
  \end{center}
\end{figure}

DBNs have a strong expressive power, and they subsume and generalise a variety
of classic models that have been studied individually in the literature. Here we
will discuss three such models: hidden Markov models, vector auto-regressive
models and Kalman filters. Hidden Markov models and Kalman filters can be seen
as particular instances of state-space models, and their DBN representations
make clear the relationship between these two classes of models. Representing
these models as DBNs has advantages beyond making their comparison more
convenient; it makes it possible to use the probabilistic machinery we covered
in Sections \ref{sec:background} and \ref{sec:dbn-intro} when more convenient or
computationally efficient than the alternatives \citep[as in the case of
inference in hidden Markov models;][]{murphy}. Some examples are shown in
Figure \ref{fig:dbns}, and will be illustrated below.

\emph{Hidden Markov models} \citep[HMMs;][]{zucchini} are one of the most
widespread approaches to model phenomena with hidden state, that is, in which
the behaviour of the observed variables $\X$ depends on that of one or more
discrete latent variables $\Z$ as well as on other variables in $\X$. This
scenario commonly arises when technical, economic or ethical considerations make
it unfeasible to completely observe the underlying state of the phenomenon of
interest. Notable examples are imputation \citep{marchini} and phasing
\citep{marchini2} in genome-wide association studies, due to limitations in the
technology to probe and tag DNA; tracking animals in ecology \citep{patterson},
where we can observe their movements using radio beacons but not their
behaviour; and confirming mass migrations through history by combining
archaeological artefacts and ancient DNA samples \cite{durbin}. Until the advent
of deep neural networks, HMMs were also the choice model for speech
\citep{gales} and handwriting recognition \citep{fink}.

In DBN terms, a typical HMM model with $M$ latent variables can be written as
\begin{align}
  &\Prob\left(\Xt \given \Xtt, \Zt\right) =
    \prod_{i = 1}^N \Prob\left(\Xit \given \PXit, \Zt\right)&
  &\text{and}&
  &\Prob\left(\Zt\right) =
     \prod_{j = 1}^{M} \Prob\left(\Zjt \given \PZjt\right),
\end{align}
with the restriction that the parents of $\Zjt$ can only be other latent
variables. Latent variables are assumed to be discrete; and in the vast majority
of the literature observed variables are assumed to be discrete as well.
Depending on the choice of $\PZjt$, we can obtain various HMM variants such
as hierarchical HMMs \citep{hhmms}, in which each $\Zjt$ is defined as an
HMM itself to produce a multi-level stochastic model; and factorial HMMs
\citep{factorial-hmms}, in which the $\Xt$ are driven by the configuration of a
set of mutually independent $\Zjt$ (shown in Figure \ref{fig:dbns}, top-left
panel).

\emph{Vector auto-regressive models} \citep[VARs;][]{box} are a straightforward
multivariate extension of univariate auto-regressive time series for continuous
variables. As such, their major applications are forecasting in finance
\citep{banbura} and more recently in the analysis of fMRI data \citep{gates}.
They are defined as
\begin{align}
  &\Xt = A_1 \Xtt + \ldots + A_L \X^{(t - L)} + \varepsilon_t,&
  &\varepsilon_t \sim N(0, \Sigma), A_1, \ldots A_L \in \RR^{N \times N},
\label{eq:vars}
\end{align}
for some fixed Markov order $L$. We can rewrite \mref{eq:vars} as
\begin{equation*}
   \Xt \given \Xtt, \ldots, \X^{(t - L)} \sim
     N\left(A_1 \Xtt + \ldots + A_L \X^{(t - L)}, \varepsilon_t\right)
\end{equation*}
and then restrict the parents of each $\Xit$ to those for which the
corresponding regression coefficients in $A_1, \ldots, A_L$ are different from
zero using the one-to-one correspondence between regression coefficients and
partial correlations \citep{weatherburn}. Formally, $X_j^{(t - l)} \in \PXit$ if
and only if $A_l[i, j] \neq 0$, which makes it possible to write \mref{eq:vars}
in a similar form to \mref{eq:dbn-def} and obtain a Gaussian DBN.  The same
construction is used by \citet{song} for their non-homogeneous DBN models, which
are parameterised as VAR processed and estimated using $L_1$-penalised
regressions. In the special case in which $L = 1$, VARs can be graphically
represented in two equivalent ways shown in the bottom panels of Figure
\ref{fig:dbns}: an ``unrolled'' DBN in which each node corresponds to a single
$\Xit$; and a more compact ``rolled-up'' DBN in which each node corresponds to a
variable $X_i$, an arc from $X_i$ to $X_j$ implies $X_i^{(t - 1)} \to
X_j^{(t)}$. An arc from $X_i$ to itself implies $X_i^{(t - 1)} \to \Xit$ as a
special case for $i = j$.

\emph{Kalman filters} \citep[KFs;][]{kalman} combine traits of both HMMs and
VARs, as discussed in depth in \citet{zoubin} and \citet{zoubin2}: like VARs,
they are linear Gaussian DBNs; but they also have latent variables like HMMs.
They are widely used for filtering (that is, denoising) and prediction in
GPS positioning systems \citep{gps}; atmospheric modelling and weather
prediction \citep{wind}; and seismology \citep{tweet}. In their simplest form
(see Figure \ref{fig:dbns}, top-right panel), KFs include a layer of one or more
latent variables that model the unobservable part of the phenomenon,
\begin{align*}
  &\Zt = A \Ztt + B \Ut + \zeta_t,&
  &\zeta_t \sim N(0, \Psi), A \in \RR^{M \times M}, B \in \RR^{M \times P}
\end{align*}
feeding into one or more observed variables
\begin{align*}
  &\Xt = C \Zt + D \Ut + \varepsilon_t,&
  &\varepsilon_t \sim N(0, \Sigma), C \in \RR^{N \times M}, D \in \RR^{N \times P}
\end{align*}
with independent Gaussian noise added in both layers. Both layers often include
additional (continuous) explanatory variables $\U$ and can also be augmented
with (discrete) switching variables to allow for different regimes as in
\citet{husmeier-nips,husmeier}. If we exclude the latter, the assumption is that
the system is jointly Gaussian: that makes it possible to frame KFs as a DBN in
the same way we did for VARs.

\section{Bayesian Networks from Incomplete Data}
\label{sec:incomplete}

The vast majority of the literature on learning BNs rests on the assumption that
$\D$ is \emph{complete}, that is, a data set in which every variable has an
observed value for each sample. However, in real-world applications we
frequently have to deal with \emph{incomplete} data; some samples will be
completely observed while others will contain missing values for some of the
variables. While it is tempting to simply impute the missing values as a
preprocessing step, it has long been known that even fairly sophisticated
techniques like hot-deck imputation are problematic in a multivariate setting
\citep{hotdeck}. Just deleting incomplete samples can also bias learning,
depending on how missing data are missing.

\begin{figure}
  \begin{center}
  \includegraphics[width=0.9\textwidth]{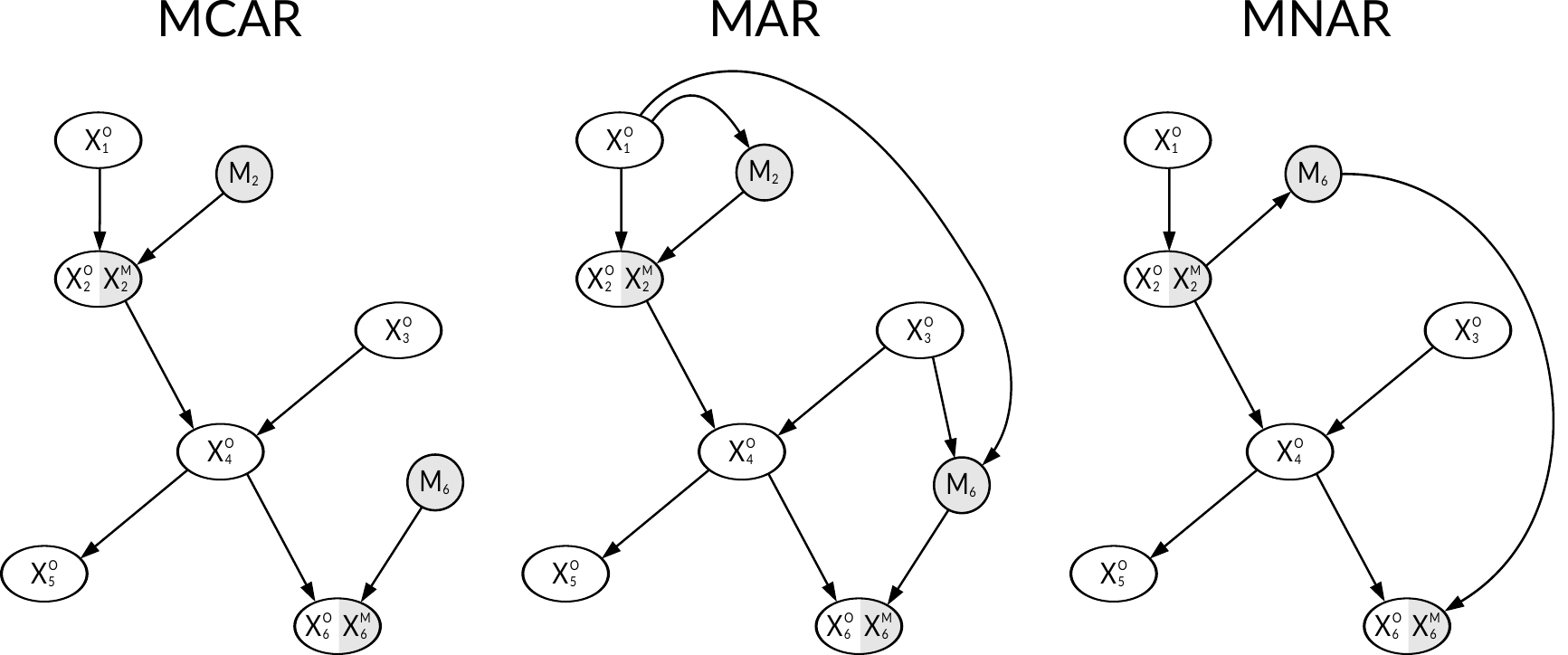}
  \caption{BN representations of the MCAR (left), MAR (centre) and MNAR (right)
    patterns of missingness from \citet{rubin2} and \citet{rubin1}. Shaded nodes
    and portion of nodes correspond to variables that are not observed in the
    data.}
  \label{fig:missing}
  \end{center}
\end{figure}

\citet{rubin2} and \citet{rubin1} formalised three possible patterns (or
\emph{mechanisms}) of missingness (illustrated in Figure \ref{fig:missing}):
\begin{itemize}
  \item Missing completely at random (MCAR): when complete samples are
    indistinguishable from incomplete ones. In other words, the probability that
    a value will be missing is independent from both observed and missing
    values. For instance, in the left panel of Figure \ref{fig:missing} both
    $X_2$ and $X_6$ are only partially observed; hence $X_2 = (X_2^O, X_2^M)$
    and $X_6 = (X_6^O, X_6^M)$ where $X_2^0$ and $X_6^O$ are observed and
    $X_2^M$ and $X_6^M$ are missing. The patterns of missingness are controlled
    by $M_2$ (for $X_2$) and $M_6$ (for $X_6$) and are completely random; say,
    $M_2$ and $M_6$ are binary variables that encode the probability of two
    instruments (independently) breaking down and thus failing to measure $X_2$
    and $X_6$ for some individuals.
  \item Missing at random (MAR): cases with incomplete data differ from cases
    with complete data, but the pattern of missingness is predictable from
    other observed variables. In other words, the probability that a value will
    be missing is a function of the observed values. An example is the central
    panel of Figure \ref{fig:missing}: compared to the left panel, we now know
    that the two instruments are likely to fail when (say) high values of $X_1$
    (and $X_3$, in the case of $X_6$) are observed.
  \item Missing not at random (MNAR): the pattern of missingness is not
    random or it is not predictable from other observed variables; the
    probability that an entry will be missing depends on both observed and
    missing values. Common examples are variables that are missing
    systematically or for which the patterns of missingness depends on the
    missing values themselves. In the right panel of Figure \ref{fig:missing},
    say that $X_2$ is censored (that is, it is never observed if its value is
    higher than a fixed threshold); but when $X_2$ is missing the probability
    that $X_6$ is missing (encoded by $M_6$) is also extremely high. Since we
    never observe $X_2$ when $X_6$ is missing, we are unable to correctly model
    this relationship; as far as we know both $X_2$ and $X_6$ may be missing due
    to some common external factor, since they appear to be missing together in
    the data.
\end{itemize}
MCAR and MAR are \emph{ignorable} patterns of missingness; the probability that
some value is missing may depend on observed values but not on missing values,
and thus can be properly modelled. If we denote with $\D^O$ and $\D^M$ the
observed and unobserved portions of $\D$, and we group all the binary
missingness indicators $M_i$ in $\M$ with parameters $\Xi$, then we can write
\begin{equation}
  \Prob\left(\D, \M \given \G,\Theta, \Xi\right) =
  \Prob\left(\D^O, \D^M, \M \given \G,\Theta, \Xi\right) =
    \int \Prob(\D^O, \D^M \given \G, \Theta) \Prob\left(\M \given \D^O, \D^M, \G, \Xi\right) \dd\D^M;
\label{eq:rubin}
\end{equation}
if the missing data are MAR then $\M$ only depends on $\D^O$,
\begin{equation*}
  \Prob\left(\M \given \D^O, \D^M, \G, \Xi\right) = \Prob\left(\M \given \D^O, \G, \Xi\right);
\end{equation*}
and if the missing data MCAR then $\M$ does not depend on either $\D^O$ or $\D^M$,
\begin{equation*}
  \Prob\left(\M \given \D^O, \D^M, \G, \Xi\right) = \Prob\left(\M \given \G, \Xi\right).
\end{equation*}
In both cases it is possible to model $\M$ from the available data. However,
this is not the case for MNAR since $\M$ depends on the unobserved $\D^M$.

Modelling incomplete data is analytically intractable and computationally
prohibitive compared to the complete data scenario: an exact analysis requires
the computation of the joint posterior distribution of $\Theta$ considering all
possible completions of $\D$. However, completing $\D$ has a computational
complexity that grows exponentially with the number of missing entries, since it
involves finding the set of missing data completions with the highest
joint probability given the observed data. Considering just the most probable
completion can also induce over-confidence in the results of the analysis, since
completed observations will have lower variability by construction; but full
Bayesian inference averaging over all possible $\Theta$ and completed $\D$ is so
computationally challenging as to be unfeasible even in simple settings.
Furthermore, this joint maximisation breaks the parameter independence
assumption and makes it impossible to define decomposable scores for structure
learning without resorting to some approximation.

\subsection{Parameter Learning}
\label{sec:paramlearn}

\begin{figure}[t]
\begin{center}
  \includegraphics[width=0.9\textwidth]{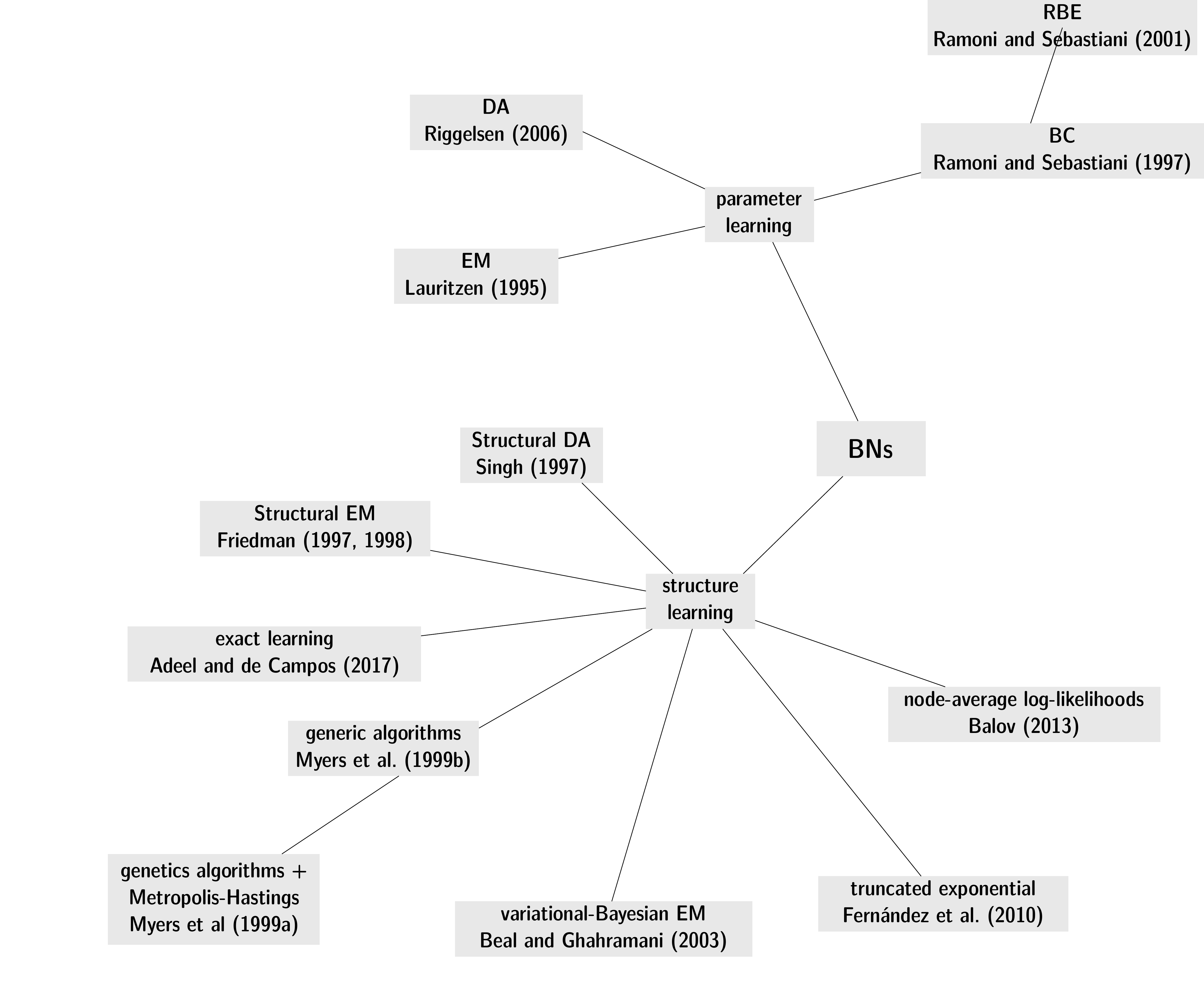}
  \caption{Diagram illustrating the relationships between the papers and the
  approaches covered in Section \ref{sec:incomplete}.}
  \label{fig:mindmap-missing}
\end{center}
\end{figure}

Many approaches have been developed for parameter learning from incomplete data
given a fixed, known structure\footnote{Although almost all the literature
specific to BNs assumes discrete data.}, ranging from iterative methods like
Data Augmentation \citep[DA;][]{data-augmentation} and the
Expectation-Maximisation algorithm \citep[EM;][]{lauritzen}, to methods based on
probability intervals such as Bound and Collapse \citep[BC;][]{sebastiani1} and
uncertain probabilities \citep{serafin}. (See Figure \ref{fig:mindmap-missing}
for an overview of those we will discuss in detail below.) These methods usually
assume missing data are MCAR or MAR to work on $\Prob(\Theta, \D^M \given \G,
\D^O)$, and their accuracy decreases dramatically if that is not the case
\citep{em-mnar}; however, BC has been found to be robust also for MNAR data.
\citet{onisko} also noted that simple imputation approaches can perform well in
learning the parameters of a BN given a fixed, sparse network structure, which
further expands available options.

In the context of parameter learning, the EM algorithm retains its classic
structure:
\begin{itemize}
  \item the \emph{expectation} (E) step consists in computing the expected
    values of the sufficient statistics (such as the counts $n_{ijk}$ in
    discrete BNs, partial correlations in GBNs), using exact inference along the
    lines described above to make use of incomplete as well as complete samples;
  \item the \emph{maximisation} (M) step takes the sufficient statistics from
    the E-step and estimates the parameters of the BN, either using maximum
    likelihood or Bayesian posterior estimators.
\end{itemize}
The parameter estimates are then used in the next E-step to update the expected
values of the sufficient statistics; repeated iterations of these two steps will
in the limit return the maximum likelihood or maximum \emph{a posteriori}
estimates for the parameters. Using the notation of \mref{eq:rubin}, the E-step
is equivalent to computing $\E(\D^M, \D^O \given \G, \Theta)$ and
the M-step to maximising $\Prob(\Theta \given \G, \D^O, \D^M)$.

DA is quite similar, but instead of converging iteratively to a single set of
parameter estimates it uses Gibbs Sampling \citep{geman} to generate values from
the posterior distributions of both $\D^M$ and $\Theta$. The two steps are as
follows:
\begin{itemize}
  \item in the \emph{imputation} (I) step  the data are completed with values
    drawn from the predictive distributions of the missing values;
  \item in the \emph{parameter} (P) estimation step a parameter value is drawn
    from the posterior distribution of $\Theta$ conditional on the completed
    data from the I-step.
\end{itemize}
More formally, we define the augmented parameter vector $\{ \D^M, \Theta \}$
containing both the missing values and the parameters of the BN. Given an
initial set of values, it updates each element of $\{ \D^M, \Theta \}$ by
sampling a new value for each missing value from $\Prob(\D^M_{i} \given
\D^M_{-i}, \Theta)$, and by sampling a new value for each parameter from
$\Prob(\Theta_{i} \given \Theta_{-i}, \D^M)$, each in turn. After an initial
burn-in phase, this process will converge to its stationary distribution and
return parameter realisations from the posterior distribution of $\Theta$
conditional on the observed data. A similar Gibbs sampling approach has been
proposed more recently by \citet{riggelsen}: it samples from a simpler,
approximate predictive distribution and makes use of weights to implement an
efficient importance sampling scheme. In addition, the weights make it possible
to use samples generated in the burn-in phase as well as those from the
stationary distribution of the Gibbs samples because they weight samples
according to their estimated predictive accuracy.

Finally, BC and its successor the Robust Bayesian Estimator
\citep[RBE;][]{sebastiani2} exploit the discrete nature of categorical variables
to produce rough interval estimates of the conditional probabilities learned
from incomplete variables, which are then reduced to point estimates either via
a convex combination of the intervals' bounds, expert knowledge or both.  The
first (\emph{bound}) step, uses the fact that each $\piijk$ can be bounded below
by assuming that none of the missing values for $X_i$ are completed with their
$k$th value when the $j$th parent configuration is observed; and that $\piijk$
can be bounded above by assuming that all missing values are completed with
their $k$th value. This approach has the merit of not making any assumptions on
the distribution of missing data. Furthermore, the width of each interval
provides an explicit representation of the reliability of the estimates, which
can be taken into account in inference and prediction. The second
(\emph{collapse}) step assumes missing data are MAR or MCAR to be able to
compute the expected completions for incomplete samples, which are then used to
compute the mean and variance of the $\piijk$. Interestingly, the intervals from
the \emph{bound} step can be used to augment both EM and Gibbs sampling and
obtain more precise inferences, but the predictive accuracy of RBE was shown to
be superior to both in \citet{sebastiani2}. A similar investigation in the
context of BN classifiers can be found in \citet{pena}. A conceptually similar
approach was also proposed by \citet{liao}, which first used qualitative expert
knowledge on the parameter values to bound them, and then estimated their values
using convex optimisation embedded in the EM algorithm.

Note that we can work on latent variables using similar approaches as long as
$\G$ is fixed and we just need to learn $\Theta$. A recent example is
given in \citet{motomura}, who show it is possible to learn the domain of a
latent discrete variable as well as the associated parameters as long as it has
observed parents and children. Several other examples are discussed in the
context of DBNs in \citet{murphy}.

\subsection{Structure Learning}
\label{sec:structlearn}

Learning the structure of a BN from incomplete data is, in many respects, an
extension of the techniques covered in Section \ref{sec:paramlearn}. (See Figure
\ref{fig:mindmap-missing} for an overview of the approaches we will discuss in
detail below as well as those for parameter learning.) In its general form
it is computationally unfeasible because we need to perform a joint optimisation
over the missing values and the parameters to score each candidate network.
Starting from \mref{eq:learning}, we can make this apparent by rewriting
$\Prob(\D \given \G)$ as a function of $\D^O, \D^M$:
\begin{multline}
  \Prob\left(\D \given \G\right) = \Prob\left(\D^O, \D^M \given \G\right)
    = \int \Prob\left(\D^O, \D^M \given \G, \Theta\right)
           \Prob\left( \Theta \given \G \right) \dd\Theta = \\
    = \int \underbrace{\Prob\left(\D^M \given \D^O, \G, \Theta\right)}_{\text{missing data}}
           \underbrace{\Prob\left(\D^O \given \G, \Theta\right)}_{\text{observed data}}
           \underbrace{\Prob\left(\Theta \given \G\right) \dd\Theta}_{\text{averaging over parameters}}.
\label{eq:marginal}
\end{multline}
From this expression we can see that in order to maximise $\Prob(\D \given \G)$
we should jointly maximise the probability of the observed data $\D^O$ and the
probability of the missing data $\D^M$ given the observed data, for each
candidate $G$ and averaging over all possible $\Theta$. This gives us the
\emph{maximum a posteriori} solution to structure learning; a full Bayesian
approach would require averaging over all the possible configurations of the
missing data as well, leading to
\begin{equation}
  \Prob\left(\D \given \G\right)
    = \iint \Prob\left(\D^M \given \D^O, \G, \Theta\right)
            \Prob\left(\D^O \given \G, \Theta\right)
            \Prob\left(\Theta \given \G\right) \dd\Theta \dd\D^M.
\label{eq:marginal2}
\end{equation}
Compared to \mref{eq:marginal}, \mref{eq:marginal2} contains one extra dimension
for each missing value (in addition to one dimension for each parameter in
$\Theta$) and thus it is too high-dimensional to compute in practical
applications. An additional problem is that, while $\Prob(\D^O \given \G, \Theta)$
decomposes as in \mref{eq:decomp}, $\Prob(\D^M \given \D^O, \G, \Theta)$ does
not in the general case.

In order to sidestep these computational issues, the literature has
pursued two possible approaches: iteratively completing and refining the data,
an using standard algorithms and scores for complete data; or using scoring
functions that approximate BIC and $\Prob(\D \given \G)$ but that are
decomposable and can be computed efficiently even on incomplete data.

The Structural EM algorithm \citep[SEM\footnote{This acronym is another source
of confusion, since SEM can also stand for ``structural equation models'' which
are closely related to BNs. See \citet{gupta} for a discussion of their
similarities and differences.};][]{sem1} is the most famous implementation of
the first approach; it has important applications in phylogenetics
\citep{sem-phylo}, clinical record \citep{sem-ecr} and clinical trial analysis
\citep{wadpain18}. SEM makes structure learning computationally feasible by
searching for the best structure inside of EM, instead of embedding EM inside a
structure learning algorithm. It consists of two steps like the classic EM:
\begin{itemize}
  \item in the E-step, we complete the data by computing the expected sufficient
    statistics using the current network structure;
  \item in the M-step, we find the structure that maximises the expected score
    function for the completed data.
\end{itemize}
Since the scoring in the M-step uses the completed data , structure learning can
be implemented efficiently using standard algorithms. The original proposal by
\citet{sem1} used BIC and greedy search; \citet{sem2} later extended SEM to a
fully Bayesian approach based posterior scores, and proved the convergence of
the resulting algorithm.

In fact, any combination of structure learning algorithm and score can be used
in the M-step; most recently \citet{scanagatta2} proposed learning BNs with a
bounded-treewidth structure using their k-MAX algorithm and a variant of BIC.
(This has the two-fold advantage of speeding up the M-step and of yielding BNs
for which completing data in the E-step is relatively fast.) \citet{singh}
proposed a similar approach based on DA, generating sets of completed data sets
and averaging the resulting learned networks in each iteration. \citet{myers}
also chose to iteratively learn both the network structures and the missing
data at the same time, but did so using evolutionary algorithms and encoding
both as ``genes''.  Hence, the individuals in the population being evolved
comprise both a completed data set and the associated BN. This was combined with
Metropolis-Hastings to speed up learning as discussed in \citet{myers2}.

More recently, \citet{cassio} proposed an exact learning algorithm that
explicitly models the patterns of missingness with auxiliary variables, which
are included as separate nodes in the BN rather than just being computational
devices. Additionally, they showed that its computational complexity is the same
as that of other exact learning algorithms for complete data; and they adapted
the proposed algorithm into a (faster) heuristic that is then proven to be
consistent.

The second group of approaches includes the variational-Bayesian EM from
\citet{zoubin4} that maximises a variational approximation of $\Prob(\D \given
\G)$, which in turn is a lower bound to the true marginal likelihood.
\citet{balov} proved that structure learning with BIC is not consistent even
under MCAR, and suggested replacing it with node-average penalised
log-likelihoods computed from locally complete observations. An alternative
consists in using approximations based on mixtures of truncated exponentials, as
was showcased in \citet{salmeron}: they combined EM and DA to fit
regression-like BNs with structures that resemble naive Bayes and tree-augmented
naive Bayes classifiers, and approximating explanatory variables.  Approachable
introductions to this area of research are provided in \citet{chickering} and
\citet{heckerman4}, which describe the relationship between $\Prob(\D \given
\G)$, its Laplace approximation and BIC in mathematical detail.

\section{Software and Applications}
\label{sec:applications}

To conclude this review, we consider the practicalities of using the BN models
discussed above to investigate real-world problems: the availability of software
implementations and notable examples of applications in various fields.

\subsection{Available Software}
\label{sec:software}

Among the DBN papers we covered, only \citet{husmeier-nips,husmeier} and
\citep{lahdesmaki} provide implementations of the methods they propose as Matlab
scripts available upon request from the authors. Additionally, \citet{stella}
provides links to the custom software they use in the paper. Hugin \citep{hugin}
implements discrete-time, homogeneous DBNs as defined in \mref{eq:dbn-def}, as
does GeNIe \citep{genie}.

Software implementations for handling incomplete are more readily available,
even though typically not from the papers that propose the methods. Both Hugin
and GeNIe implement EM for parameter learning, but not Structural EM for
parameter learning. The R packages bnlearn \citep{jss09} and bnstruct
\citep{bnstruct} implement the structural EM algorithm; another implementation
in Matlab is made available as part of the supplementary material of
\citep{cassio-sem}. \citet{cassio} have later used it to implement their
exact learning algorithm along with Gobnilp \citep{gobnilp}; and
\citep{riggelsen} uses a custom C++ implementation of his Gibbs sampling
approach. The approaches proposed by \citet{balov} are implemented in
the catnet R package \citep{catnet}, and those proposed by \citet{salmeron}
are included in Elvira \citep{elvira}.

\subsection{Notable Applications}
\label{sec:notable}

Dynamic and incomplete data are common in many application fields in which BNs
are used; the need to work with such data is the main motivation for developing
the approaches we covered in Sections \ref{sec:dynamic} and
\ref{sec:incomplete}. Therefore, many of them have been used in advanced
applications especially in the clinical and life sciences.

This is especially true for DBNs: since they subsume a number of statistical
models that are themselves very popular (see Section \ref{sec:special-cases}),
it would be fair to say that they are used in most branches of science under
different guises. However, their main application fields in the literature
remain genetics and systems biology, and in particular the modelling of cell
signalling pathways in the form of networks as they evolve over time. This
involves probing the state of multiple proteins or other omics simultaneously
through time, and studying their interactions and how they evolve in response to
external stimuli. \emph{In vitro} experiments, in which cell lines are probed in
a controlled environment, have been modelled as homogeneous DBNs \citep[see, for
instance,][]{cancer-dbn,perrin}.  On the other hand, more complex and \emph{in vivo}
experiments often involve molecular systems that are not in a steady-state
regime and thus require non-homogeneous DBNs. Some examples based on the models
presented in \citet{husmeier-nips,husmeier}, \citet{husmeier-latest} and other
papers from the same authors study embryonic stem cells \citep{stem-cells};
sepsis as a cause of acute lung injury \citep{sepsis}; blunt trauma and
traumatic spinal cord injury association with inflammation \citep{trauma} and
hypotension \citep{trauma2}. Other omics data including transcriptomics
\citep{transcriptomics} and microbiome data \citep{microbiome2,microbiome} have
been studied using DBNs as well.

DBNs have also seen application in ecology, specifically in studying species
dynamics. Like the human body, an ecosystem acts as a complex system where
species interact with each other (for example, through predation or competition)
and with the environment they live in. The resulting dynamics have been modelled
in \citet{eco-dbn} with the non-homogeneous DBN from \citet{husmeier}. A DBN
that corrects for spatial correlation as well through the use of latent
variables have been developed in \citet{fish-dbn} to study fish species dynamics
in 7 geographically and temporally varied areas within the North Sea.
Furthermore, ecosystems have been studied using DBNs from the point of view of
environmental sciences as well, considering for example the impact of climate
change on groundwater \citep{groundwater} and how to best manage water
reservoirs under infrequent rainfalls \citep{reservoirs}.

Finally, a third class of applications of DBNs is modelling mechanical systems
in engineering, reliability and quality control studies. For instance, studying
the deterioration of steel structures subject to fatigue \citep{fatigue}
including warships \citep{warships}; implementing real-time reliability
monitoring of subsea pipes \citep{pipes} and CNC industrial systems \citep{cnc};
fault detection, identification and recovery in autonomous spacecrafts
\cite{mars}; and predicting traffic dynamics using probes in San Francisco
\citep{artery}  In each of these applications, DBNs capture the interactions of the
components of the mechanical systems they are modelling while at the same time
incorporating the effects of the surrounding environment, thus allowing the
operators (or the systems itself) to adjust its mode of operation or to stop
it completely before causing any damage or safety issue. Software systems have
been similarly modelled, for example to study the relationship between security
vulnerabilities and attack vectors in network security \citep{security}.

As for incomplete data, they are such a widespread issue in science and
engineering that it is difficult to identify any particular field in which they
are prevalent; hence the BN learning approaches in Section \ref{sec:incomplete}
have found very different applications in the literature. We note, however,
that the analysis of clinical data has historically been one of the key
applications that has driven methodological developments in this area. We
mentioned earlier the study of clinical records of patients chronic obstructive
pulmonary disease \citep{sem-ecr} and the recovery process of patients with
whiplash-associated disorders \citep{wadpain18}; two other examples are the
diagnosis of brain diseases such as dementia and Alzheimer \citep{alzheimer}
and of sleep apnea \citep{apnea}. Completely different applications have also
been explored: one example is the classification of radar images \citep{radar}.

\section{Summary}

In this paper we have reviewed the fundamental definitions and properties of
BNs, and how BNs can be stretched to encode more complex probabilistic models
than what might be apparent from reference material and most of the literature.
Both have an overwhelming focus on the straightforward case in which the data
being modelled are both static (that is, with no time dimension) and complete
(that is, with no missing values). However, dynamic and incomplete data are
central to many cutting-edge applications in research fields ranging from
genetics to robotics; BNs can play an important role in many of these settings,
as has been evidenced by the examples referenced in Sections \ref{sec:dynamic}
and \ref{sec:incomplete}. Given their expressive power, BNs also subsume several
classic probabilistic models and can augment them with automatic reasoning
capabilities through various kinds of queries that can be performed
algorithmically. Much research has been and is being developed to adapt BNs to
these applications and make them a competitive choice for modelling complex
data.

\selectlanguage{english}
\clearpage

\end{document}